\documentclass[sn-nature,Numbered]{sn-jnl}

\usepackage{setspace}
\usepackage{graphicx}%
\usepackage{multirow}%
\usepackage{amsmath,amssymb,amsfonts}%
\usepackage{amsthm}%
\usepackage{mathrsfs}%
\usepackage[title]{appendix}%
\usepackage{xcolor}%
\usepackage{textcomp}%
\usepackage{manyfoot}%
\usepackage{booktabs}%
\usepackage{algorithm}%
\usepackage{algorithmicx}%
\usepackage{algpseudocode}%
\usepackage{listings}%
\theoremstyle{thmstyleone}%
\theoremstyle{thmstyletwo}%

\theoremstyle{thmstylethree}%

\begin{document}

\title[Article Title]{Resolved ALMA observations of the water content of the inner astronomical units of HL Tau}

\author*[1]{\fnm{Stefano} \sur{Facchini}}\email{stefano.facchini@unimi.it}

\author[2]{\fnm{Leonardo} \sur{Testi}}

\author[3,4,5]{\fnm{Elizabeth} \sur{Humphreys}}

\author[6]{\fnm{Mathieu} \sur{Vander Donckt}}

\author[7]{\fnm{Andrea} \sur{Isella}}

\author[7]{\fnm{Ramon} \sur{Wrzosek}}

\author[8]{\fnm{Alain} \sur{Baudry}}

\author[9]{\fnm{Malcom D.} \sur{Gray}}

\author[10]{\fnm{Anita M. S.} \sur{Richards}}

\author[11]{\fnm{Wouter} \sur{Vlemmings}}

\affil*[1]{\orgdiv{Dipartimento di Fisica}, \orgname{Universit\`a degli Studi di Milano}, \orgaddress{\street{via Celoria 16}, \city{Milano}, 
\country{Italy}}}

\affil[2]{\orgdiv{Dipartimento di Fisica e Astronomia ``Augusto Righi''}, \orgname{Università di Bologna}, \orgaddress{\street{Viale Berti
Pichat 6/2}, \city{Bologna}, \country{Italy}}}

\affil[3]{ \orgname{European Southern Observatory}, \orgaddress{\street{Karl-Schwarzschild-Strasse 2}, \city{Garching bei M\"unchen}, \postcode{D-85748}, \country{Germany}}}

\affil[4]{ \orgname{Joint ALMA Observatory}, \orgaddress{\street{Alonso de Cordova 3107}, \city{Santiago},  \country{Chile}}}

\affil[5]{ \orgname{European Southern Observatory (ESO) Vitacura}, \orgaddress{\street{Alonso de Cordova 3107}, \city{Santiago},  \country{Chile}}}

\affil[6]{ \orgdiv{Space sciences, Technologies \& Astrophysics Research (STAR) Institute},\orgname{University of Li\`ege}, \orgaddress{\city{Li\`ege},  \country{Belgium}}}

\affil[7]{\orgdiv{Department of Physics and Astronomy}, \orgname{Rice University}, \orgaddress{\street{6100 Main Street, MS-108}, \city{Houston}, \postcode{77005}, \state{TX},
\country{USA}}}

\affil[8]{\orgdiv{Laboratoire d'Astrophysique de Bordeaux}, \orgname{Univ. de Bordeaux, CNRS}, \orgaddress{\street{B18N, all\'ee Geoffroy Saint-Hilaire}, \city{Pessac}, \postcode{33615}, 
\country{France}}}

\affil[9]{ \orgname{National Astronomical Research Institute of Thailand}, \orgaddress{\street{260 Moo 4, T. Donkaew, A. Maerim}, \city{Chiangmai}, \postcode{50180}, 
\country{Thailand}}}

\affil[10]{\orgdiv{JBCA},  \orgname{University of Manchester}, \orgaddress{ \postcode{M13 9PL}, 
\country{UK}}}

\affil[11]{\orgdiv{Department of Space, Earth and Environment}, \orgname{Chalmers University of Technology}, \orgaddress{\city{G\"oteborg}, \postcode{SE-412 96}, 
\country{Sweden}}}


\abstract{The water molecule is a key ingredient in the formation of planetary systems, with the water snowline being a favorable location for the growth of massive planetary cores. Here we present ALMA data of the ringed protoplanetary disk orbiting the young star HL Tauri that show centrally peaked, bright emission from water vapour in three distinct transitions of the main water isotopologue. The spatially and spectrally resolved water content probes gas in a thermal range down to the water sublimation temperature. Our analysis implies a stringent lower limit of 3.7 Earth oceans of water vapour available within the inner 17 astronomical units. We show that our observations probe the water content in the atmosphere of the disk, due to the high dust column density and absorption, and indicate the main water isotopologue as the best tracer to spatially resolve water vapour in protoplanetary disks.}

\maketitle

The water molecule is undoubtedly one of the most important molecular species in the whole universe. Being an extremely efficient solvent, water had a key role in the emergence of life as we know it on our planet. For this reason, the chemical characterization of exoplanetary atmospheres is often focused on detecting this particular molecular species \cite{Kreiberg_ea_2014,madhu19,Rustamkulov2023}. Formed by the common H and O atoms, water plays a fundamental role in the physics of the formation of planetary systems, due to its very high abundance in both gaseous and icy forms \cite{vanDishoeck2014,Drazkowska2023}. Theoretical models predict that at the location of the phase transition from gaseous to solid form, dust grains can accumulate and grow very efficiently, promoting the fast formation of  planetary cores.  Across this particular radial location, called `snowline', grains can drastically change their drift and fragmentation velocity, composition, and opacity. In synergy with vapour radial diffusion \cite{Cuzzi_ea_2004}, these physical discontinuities can lead to the accumulation and  growth of dust grains into planetesimals \cite{Schoonenberg_Ormel_2017,Drazkowska_Alibert2017}. The position of the snowline also defines
the chemistry of the available planet building blocks. Since the H$_2$O molecule is the major elemental
oxygen carrier in the disk, its desorption and freezing affect the elemental C/O ratio in both the gas and
solid phases \cite{oberg11,Eistrup2016,Oberg2023}.

Because of its large binding energy, the H\textsubscript{2}O transition from ice to gas happens a few astronomical units (au) from the young star where the midplane temperatures are in the range from 100 to 200 K, making it the last major ice component to sublimate. However, the proximity to the host star makes the detection of the snowline complicated even in the closest star forming regions. Both cold and warm water lines have been detected in a few disks by {\it Herschel} (see \cite{Hogerheijde2011,van_dishoeck_ea_2021} and references therein), {\it Spitzer} \cite{Pontoppidan_ea_2010}, JWST \cite{Grant2023,Kospal2023,Banzatti2023} and ground based observatories \cite{Salyk_ea_2015}, but the low angular resolution did not allow robust inferences about the extent of the water snowline. Observing directly water emission from the ground is complicated by the high water vapour content of the Earth's atmosphere, resulting in strong telluric absorption. To circumvent this problem, most programs at mm wavelengths have focused on attempting the detection of the rarer H$_2^{18}$O and HDO isotopologues, leading to the clear detection of spatially resolved water isotopologue emission in the outbursting source V883 Ori \cite{Tobin2023}. In quiescent sources, except the candidate detections in the AS 205N and HL Tau disks, no detection of thermal emission at (sub-mm) wavelengths has been reported in the literature \cite{Kristensen_ea_2016,Carr2018,Bosman_ea_2021,Notsu2019}.

In this paper we focus on the text-book case of HL Tau, the first protoplanetary disk imaged at very high angular resolution ($\sim0.025''$) with the Atacama Large Millimeter/submillimeter Array (ALMA) \cite{HL_Tau_SV_2015}. The disk shows a spectacular pattern of concentric rings. With the source being young ($\lesssim1\,$Myr), and the dynamical stellar mass being relatively high ($2.1\pm0.2\,M_{\odot}$, \cite{Yen2019}), the inner disk temperatures are expected to be warm, due to irradiation and accretion heating. Warm and hot water has been detected in HL~Tau both by {\it Herschel} in the Far Infrared (FIR, \cite{Riviere2012}) and by ground-based high resolution spectroscopy in the Mid-Infrared (MIR, \cite{Salyk2019}), with the lines not being spatially resolved. With this article, we report the detection of three rotational water lines in the inner regions of the HL Tau protoplanetary disk obtained with ALMA. The lines are spectrally resolved. Analysis of the interferometric data confirms that the extent of the water emission is confined within a prominent gap seen in the HL~Tau continuum intensity. These new data present spatially resolved images of the emission from the main water isotopologue (H$_2$O) in a protoplanetary disk, and pave the way to a new observational strategy to characterize the water vapour content of terrestrial planet forming regions. 

\section*{Observations}
We observed HL Tau in two different ALMA bands (Band~5, originally developed with the goal of studying water in the local Universe \cite{Belitsky2018}, and Band~7), to target three transitions of water (two lines of para-water, and one line of ortho-water): p-H$_2$O $3_{13}-2_{20}$ and $5_{15}-4_{22}$, at 183.31 and 325.15\,GHz, respectively, and o-H$_2$O $10_{29}-9_{36}$ at 321.22\,GHz. The first two lines are expected to trace warm water vapour outside the water snowline, while the third line is predicted to detect hot water within the water snowline \cite{Notsu2017,Notsu2018}. We also observed a rotational transition of the water isotopologue p-H$_2^{18}$O at 322.46\,GHz. The molecular coefficients of the lines, and sensitivity of the observations, are reported in Tables~1--2.

After self-calibration, we imaged both the continuum and the continuum-subtracted lines with the CASA software \cite{casa_software}. The continuum images are shown in Figure~1. All the H$_2$O lines and the H$_2^{18}$O line were imaged with {\tt CASA} 6.2.1 with natural weighting, to maximise point source sensitivity.  The 183\,GHz line presents a synthesized beam of $0.500''\times0.442''$ (PA $-3.0$\,deg), and was imaged with a channel width of $0.8\,$km\,s$^{-1}$. The resulting beam for the 325\,GHz water line is $0.640''\times0.491''$ (PA $-42.1$\,deg), with a 1\,km\,s$^{-1}$ channel spacing. For the high excitation 321\,GHz line, only one long baselines execution block was available, and the beam is $0.067''\times0.061''$ (PA $12.2$\,deg), with a 5\,km\,s$^{-1}$ channel spacing. The H$_2^{18}$O line was imaged with several different channel spacings. To have a one-to-one comparison with the main isotopologue line, in the paper we show the results with a channel width of 1\,km\,s$^{-1}$. The resulting beam is $0.779''\times0.626''$ (PA $-46.7$\,deg).

The spectrum of the two lowest excitation lines was extracted over a circular area with $0.7''$ radius and is shown in Figure~2. Both the 183\,GHz and the 325\,GHz lines are clearly detected across multiple channels, with the lines being centered on the systemic velocity of the system (7.1\,km\,s$^{-1}$; \cite{Garufi_ea_2021,Garufi_ea_2022}). The 325\,GHz line shows an absorption component at $\sim10\,$km\,s$^{-1}$, as seen for other lines at similar upper energy levels \cite{Garufi_ea_2022}. The 183\,GHz line shows a width of $\sim12\,$km\,s$^{-1}$, while the higher frequency lines show a slightly broader emission. The 183\,GHz line spectrum exhibits a peak signal-to-noise ratio (snr) of $\sim9.6$, with an rms of $13.2\,$mJy over the flux density extraction area. The 325\,GHz line spectrum instead shows a peak snr of $\sim5.8$, with an rms of $14.0\,$mJy. The higher energy line at 321\,GHz does not show a detection when integrating over the same area, due to the much smaller beam and the resulting high rms. We thus extracted a spectrum over a circular area with radius $0.06''$. The peak snr is $\sim4.1$, with an rms of $3.0\,$mJy. This transition shows the broadest line profile among the three detected lines, consistent with originating from the inner regions with higher Keplerian velocities. In all cases, the integrated intensity map shows a strong detection with a peak co-located with the dust continuum peak (see Figure~1). We obtain a peak snr of 21.4, 19.8 and 8.1 in the integrated intensity maps of the 183, 325 and 321\,GHz lines, respectively.  The line fluxes extracted over a circular area centered on the continuum peak are reported in Table~1.

The intensity weighted velocity maps (moment one maps) of the 183 and 325\,GHz lines are shown in Figure~1. In the high snr map of the 183\,GHz line, disk rotation is clearly detected, with position angle and systemic velocity in agreement with other brighter lines from the same dataset \cite{Garufi_ea_2022}, indicating that a displacement of the photocenter of blue-shifted and red-shifted channels is detected at the current resolution.

For the H$_2^{18}$O line in Band 7, we extracted the spectrum with the same methodology described for the three main isotopologue lines, over a $0.7''$ radius circular area (Extended Data Figure~5). A moment 0 map was computed over the same spectral range as the main isotopologue line. The line was not detected.

\section*{The spatial distribution of water vapour}

The integrated intensity morphology of the 183 and 321\,GHz lines are remarkably different (see Figure~1), but so are the angular resolutions of the two moment maps. In order to derive radial profiles of the respective integrated intensities, we used two different approaches. First, we focused on the highest snr line of the sample (at 183\,GHz), which is also the coldest and therefore expected to trace the largest spatial extent. We averaged in frequency the interferometric data of the continuum-subtracted line in the same frequency range used to compute the moment 0 map. We then fitted the line integrated intensity visibilities with a simple Gaussian model using the {\tt galario} package \cite{galario}, after fixing the inclination and position angle to $46.7\deg$ and $138.0\deg$, respectively, as derived from high angular resolution continuum observations \cite{HL_Tau_SV_2015}. The fit  converges well to a Gaussian with $\sigma_{\rm G}=0.12\pm0.01''$ (see Figure~3 and Extended Data Figure~6;  $\sigma_{\rm G}$ is the standard deviation of the Gaussian function). At a distance of 140\,pc, this corresponds to $\sigma_{\rm G}=16.8\pm1.4\,$au. For the 321\,GHz line, the much higher angular resolution allowed us to compute the radial profile of the integrated intensity by azimuthally averaging the moment zero map, after de-projecting it by the known inclination (using the {\tt GoFish}  package \cite{gofish}).

Figure~3 compares the two integrated intensity profiles with the Band 7 azimuthally averaged continuum intensity profile. Both line profiles are clearly centrally peaked, indicating that the water vapour emission must originate above the optically thick continuum from the disk midplane. The lower excitation line is significantly more extended, showing that the water emission has a radially decreasing temperature (excitation), and that the high excitation line is not optically thick outside the central beam. The same line shows detectable emission out to $0.3''$ when boosting the snr by azimuthally averaging. The 183\,GHz temperature gradient is confirmed by fitting the high snr spectrum with a Keplerian disk model with a brightness temperature gradient, assuming that the line is close to being optically thick (Extended Data Figure~7). Declining temperature profiles are preferred to flat ones, in agreement with the derived integrated intensity profile.

The warm brightness temperature profile of the 183\,GHz line, which far exceeds the midplane temperatures obtained by analysis of multiwavelength continuum data \cite{Carrasco_ea_2019}, indicates that the water vapour we are tracing originates in the warm disk upper layers. This is further supported by the peak of the 321\,GHz emission, which is slightly shifted from the continuum peak (see Figure~1, bottom right panel). Even though the two are consistent within the astrometric precision of the data, the apparent shift agrees with tracing water vapour on the side of the disk closer to the observer, which is in the North-East \cite{Yen2019}, unocculted by the optically thick dust continuum. These findings set a stringent upper limit on the radius of the water snowline at 17\,au. A more accurate determination will require simultaneous forward-modelling of the radial and spectral profiles of all three (sub-)mm lines with the aid of thermo-chemical codes.

\section*{Water column density}

From the three main water lines, we computed the rotational diagram, under the assumption of optically thin emission and uniform excitation temperature across the energy range. While the very inner regions of the line emission are likely opaque, the bulk of the emission from the three lines cannot be optically thick, since this would imply a flux ratio that is equal to the square of the frequency ratio (in Rayleigh–Jeans regime and Local Thermodynamical Equilibrium -- LTE), which we do not observe. While masing cannot be excluded to partly contribute to the emission, in the high spectral resolution spectra of the 183 and 325\,GHz lines we have not identified individual narrow spectral features which are in general a good signature for maser action together with high flux density. From the high densities of the inner disk, masing is not expected for the three lines analyzed here, and the only contribution could originate in the very upper layers at low volume densities where collisional quenching of the masing action is less probable. In the rotational diagram, we used degeneracy quantum numbers that account for a 3:1 ortho-to-para ratio, and a partition function that considers all states within the same assumption \cite{Polyansky2018}. No rescaling of the ortho- and para- line fluxes is needed to compute the total water column. We accounted for 10\% absolute flux systematic uncertainties in the line fluxes.

The rotational diagram does not provide a unique solution for the rotation temperature, as shown in Figure~4. The Monte Carlo Markov Chain (MCMC) exploration individuates two distinct temperatures, which
realistically indicate a continuous gradient in the excitation temperature of the water vapour.
Fitting either the two lower energy lines or the two higher energy lines separately, the rotational
diagram indicates excitation temperatures of $214_{-29}^{+42}$ and $789_{-110}^{+127}$\,K, respectively,
in line with the two classes of solutions obtained in the joint fit (see Extended Data Figure~8). The lower temperature solution is sensitive to colder water vapour in the range of expected desorption temperatures of water ice in space, suggesting that the bulk of the water emission from the 183\,GHz line traces water gas in the proximity of the snow surface. The higher temperature solution is driven by warm gas in the upper layers of the terrestrial planet forming regions of the disk, which are well imaged by the high resolution 321 GHz line integrated intensity map, and likely by the lines being close to be optically thick. The excitation temperature of the warm gas is in broad agreement with MIR line fluxes in the 12.37 -- 12.41$\,\mu$m range on the same source (in particular the o-H$_2$O $16_{4\,13}-15_{1\,14}$ line with $E_{\rm up}=4948\,$K \cite{Salyk2019}). These high temperatures allow for water vapour formed in situ via gas-phase reactions \cite{Bethell2009,Baulch1972}.

All the rotational diagrams robustly constrain the column density of water gas (in the optically thin limit and above the optically thick continuuum). The joint fit shows a total water column density $\log_{10}{(N_{\rm thin}/{\rm cm}^2)} = 16.41^{+0.06}_{-0.09}$ within $0.7''$ ($\sim100\,$au) from the star. This corresponds to $\sim3.7$ Earth oceans ($7.1\times10^{-2}$ lunar masses) of water vapour. Given the optically thin assumption, this has to be considered as a tight lower limit. Since most of the emission originates from $\lesssim0.12''$ ($17\,$au), assuming that the entirety of it is confined within this radial range we obtain an averaged column density of $\log_{10}{(N_{\rm thin}/{\rm cm}^2)} \sim18.10^{+0.06}_{-0.09}$, when accounting for the disk inclination.   

The non-detection of the H$_2^{18}$O line can determine an upper limit of the average optical depth expected for the H$_2$O 325\,GHz line ($\tau_{\rm{H}_{2}\rm{O}}$). Assuming that $\tau_{\rm{H}_{2}\rm{O}}=530\,\tau_{\rm{H}_{2}^{18}\rm{O}}$ (using the oxygen isotope ratio measured in the solar wind \cite{McKeegan11}), we obtain that $\tau_{\rm{H}_{2}\rm{O}}<14$. By turning the argument around, the optically thin assumption for the 325\,GHz line provides an upper limit of $N$(H$_{2}$O)/$N$(H$_{2}^{18}$O)$=40$ in HL~Tau, well in agreement with oxygen fractionation levels in our own solar system  \cite{Schroeder2019}. The non detection of H$_{2}^{18}$O shows that observational campaigns aimed at targeting water in inner disks of quiescent disks with ALMA should privilege the main isotopologue as a first choice, with follow-up observations of robust detections targetting HDO and H$_2^{18}$O  to derive more accurate column densities and set stringent limits on the water deuteration \cite{Tobin2023}.

Assuming that the bulk of the water emission originates from $\lesssim17\,$au from the star, we can compare the total mass of water ($M_{{\rm H}_2{\rm O}}$) to the mass of dust $M_{\rm dust}$ estimated from multi-wavelength continuum analysis of ALMA and VLA data \cite{Carrasco_ea_2019}. By using the dust surface density from this study, we obtain $M_{\rm dust}\sim13M_{\rm Earth}$, which leads to a water-to-dust mass ratio of $M_{{\rm H}_2{\rm O}}/M_{\rm dust}\sim6\times10^{-5}$. This number is much lower than the expected water abundance in inner disks (water-to-dust mass ratio $\sim10^{-2}$). The optical thickness of water can only marginally alleviate the problem, given the non-detection of H$_{2}^{18}$O. Continuum subtraction could also marginally reduce the water brightness temperature \cite{Weaver2018}, but in this case is not expected to reduce the water fluxes by more than a factor of 2 (see Figure~3). The low water-to-dust ratio further strengthens the interpretation that with ALMA we are probing only the upper layers of the disk, above the optically thick screen of the dust continuum emission, which shows optical depths between $\sim5-10$ at 0.9\,mm within the inner 17\,au \cite{Carrasco_ea_2019}. The large cavity seen in the HDO and H$_2^{18}$O integrated intensity maps of V883 Ori \cite{Tobin2023} further supports that the observation of a large column of water is hindered by optically thick continuum in the inner disk.

\section*{Conclusions}

These new ALMA data reveal high significance detections of three distinct rotational transitions of the main isotopologue of water vapour in the inner regions of the ringed HL~Tau disk. These observations pave the way to the characterization of the water content of the inner regions of protoplanetary disks. The tremendous angular resolution and sensitivity of the ALMA telescope, even in spectral ranges of low atmospheric transmission, are providing spatially and spectrally resolved images of the vapour of the main water isotopologue in a planet forming disk. Analysis of the morphology of the water emission, of the spectrum of the highest snr line, and of the excitation conditions, jointly indicate that the (sub-)mm lines are probing warm gas in the disk upper layers above the water snow surface, with a radially decreasing temperature profile. The non detection of H$_2^{18}$O and the low water-to-dust ratio in the inner $17\,$au show that the observations are probing marginally optically thick gas above the opaque dust continuum emission. These results highlight that the water content of quiescent protoplanetary disks at (sub-)mm wavelengths is most efficiently unveiled by targeting the main isotopologue, in particular in disks with high continuum optical depths within the water snowline. 

\clearpage

\begin{figure*}
\begin{center}
\includegraphics[width=0.75\textwidth]{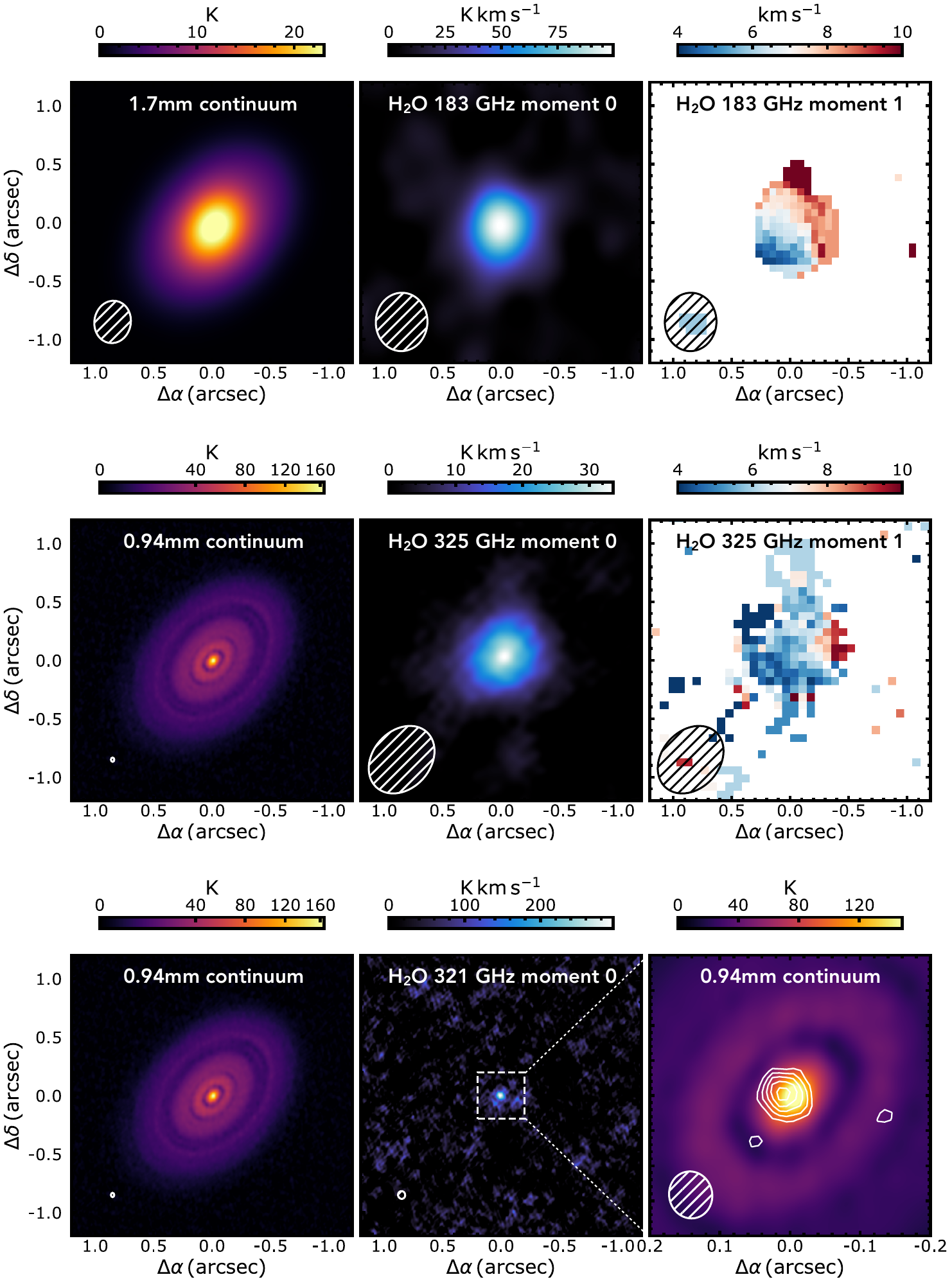}
\end{center}
\caption{{\it Top}. Left: 1.7\,mm continuum image of HL Tau. Center: integrated intensity map of the 183\,GHz water line. Right: intensity weighted velocity map of the 183\,GHz water line, after a 4$\sigma$ clipping on individual channels, where disk rotation is clearly detected. {\it Center}. Same as top panels, for the 0.94\,mm continuum and the 325\,GHz water line. The intensity weighted velocity map in this case is computed after a 3$\sigma$ clipping. {\it Bottom}. Left and center: same as top panels, for the 0.94\,mm continuum and the 321\,GHz water line. No moment one map is shown, due to low snr. Right: zoom-in of continuum intensity, with [4,5,6,7,8]$\sigma$ contours of the 321\,GHz line moment 0 map, with $\sigma=13.3\,$mJy\,beam$^{-1}$\,km\,s$^{-1}$. The rms associated to the the integrated intensity maps of the 183 and 325\,GHz lines are respectively: 28.2 and 46.3\,mJy\,beam$^{-1}$\,km\,s$^{-1}$.}
\label{fig:map}
\end{figure*}

\begin{figure*}
\begin{center}
\includegraphics[width=0.48\columnwidth]{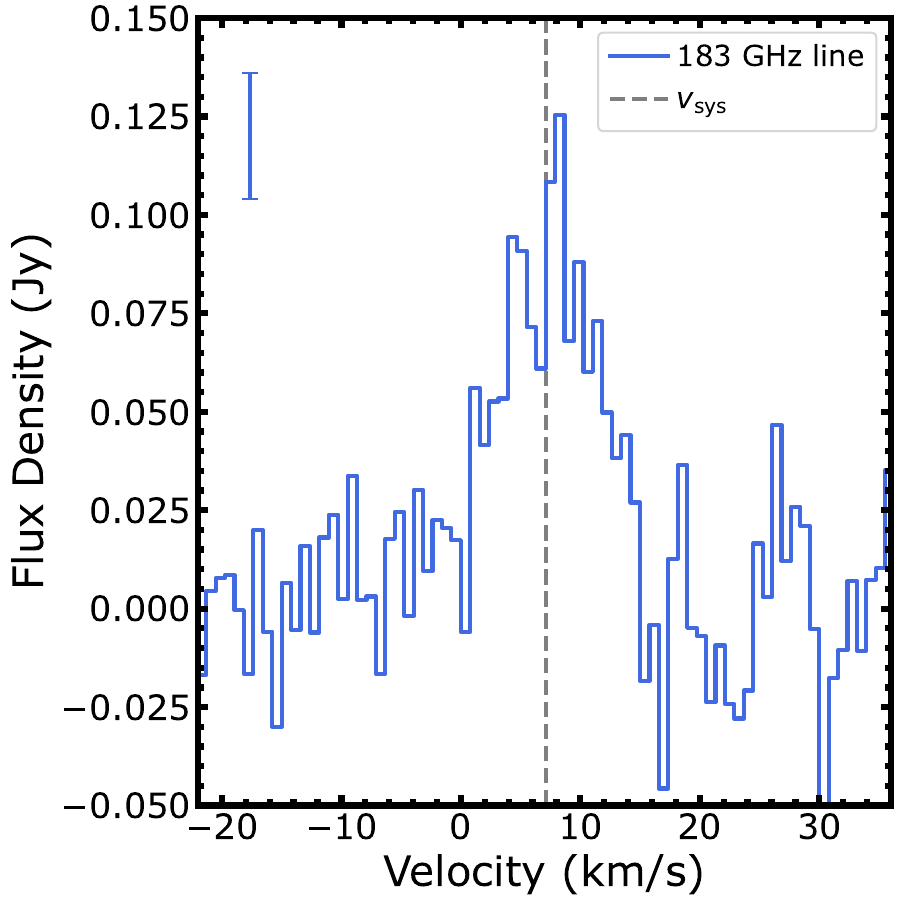}
\includegraphics[width=0.48\columnwidth]{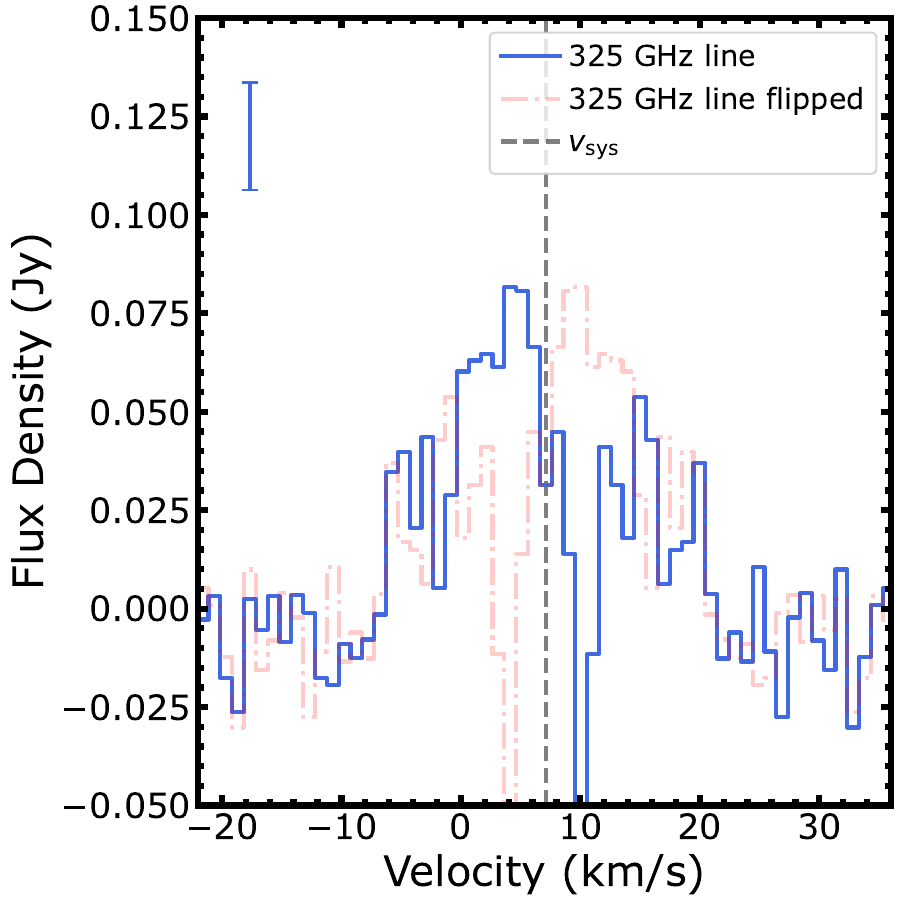}
\includegraphics[width=0.48\columnwidth]{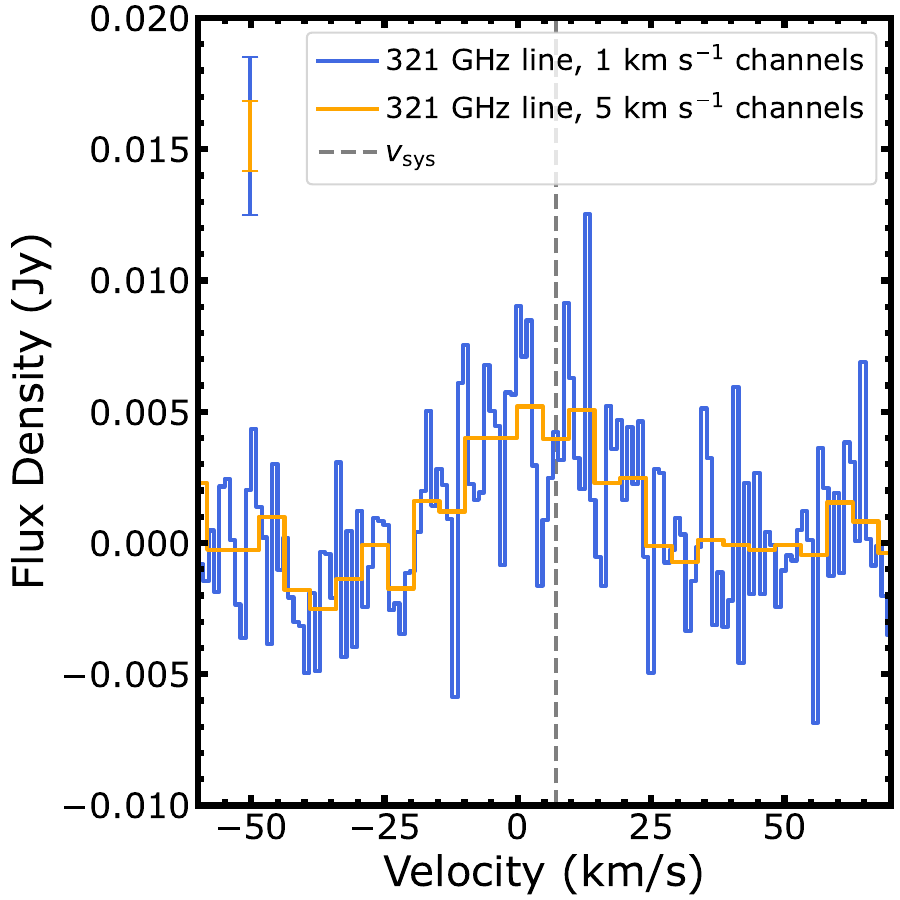}
\end{center}
\caption{Top left: spectrum of the 183\,GHz water line, extracted over a circle with radius of $0.7''$ centered on the continuum peak. Top right: spectrum of the 325\,GHz water line, extracted over the same area, highlighting the mirrored (flipped) version of the spectrum with the dashed-dotted line. Bottom: spectrum of the 321\,GHz water line extracted over a circle with radius of $0.06''$ centered on the continuum peak. The velocity range on the $x-$axis is different in the bottom panel. The grey dashed line in all panels shows the systemic velocity of $7.1\,$km\,s$^{-1}$. In the top left of each panel 2$\sigma$ scale bars are reported for reference.}
\label{fig:spectrum}
\end{figure*}

\begin{figure}
\begin{center}
\includegraphics[width=\columnwidth]
{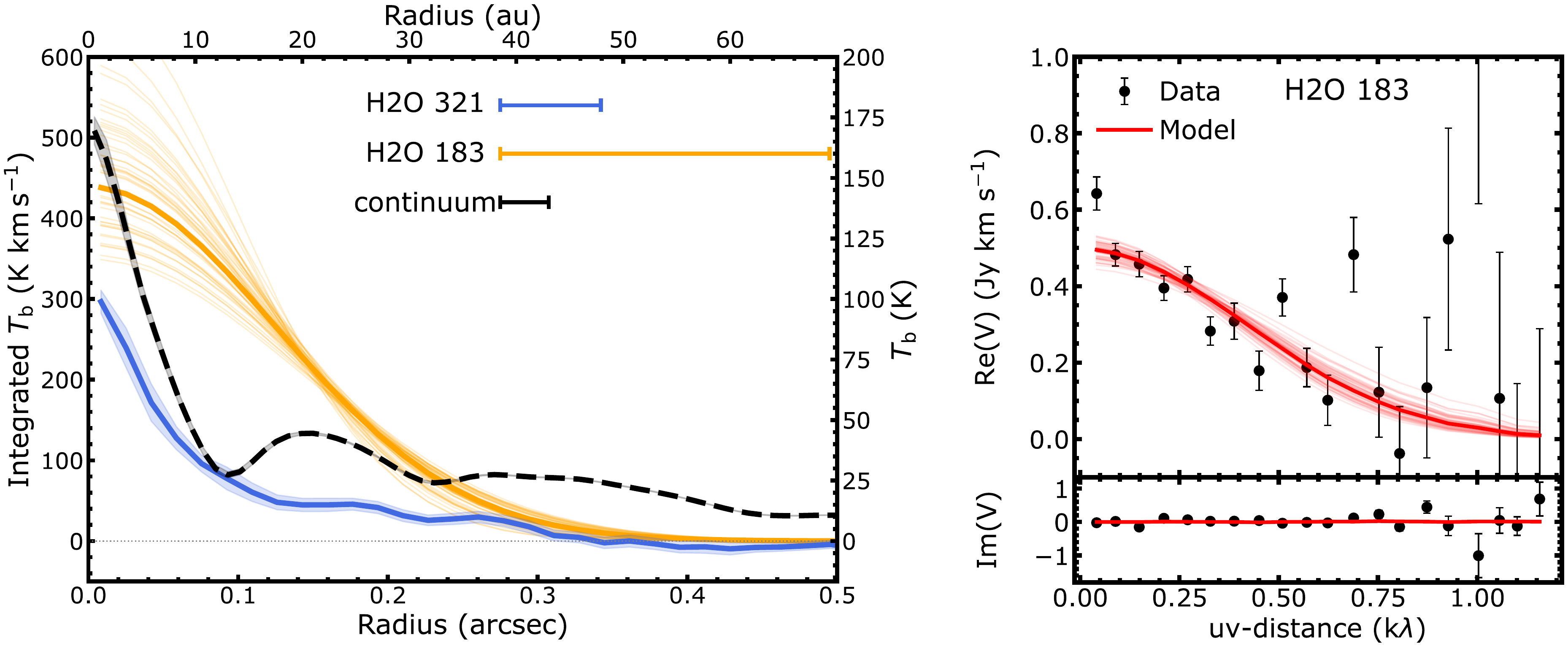}
\end{center}
\caption{Left: integrated brightness temperature ($T_{\rm b}$) radial profile of the 321\,GHz line, reconstructed integrated $T_{\rm b}$ radial profile of the 183\,GHz line, and $T_{\rm b}$ profile of the 0.94\,mm continuum emission. The thick orange line shows the best fit model assuming a Gaussian integrated intensity profile. Thin lines show randomly sampled realizations of the posterior distribution. The lines in the top right show the beam major axis. For the 183\,GHz line it portrays the smallest spatial scales to which the $uv$-plane analysis is sensitive, which are $\sim2.3$ smaller than the major axis of respective natural beam. Right: Re-centered and de-projected visibilities of the integrated intensity of the 183\,GHz water line. The thick red line shows the best fit model reproducing the profile shown on the left panel.}
\label{fig:uvplot}
\end{figure}

\begin{figure}
\begin{center}
\includegraphics[width=0.98\columnwidth]{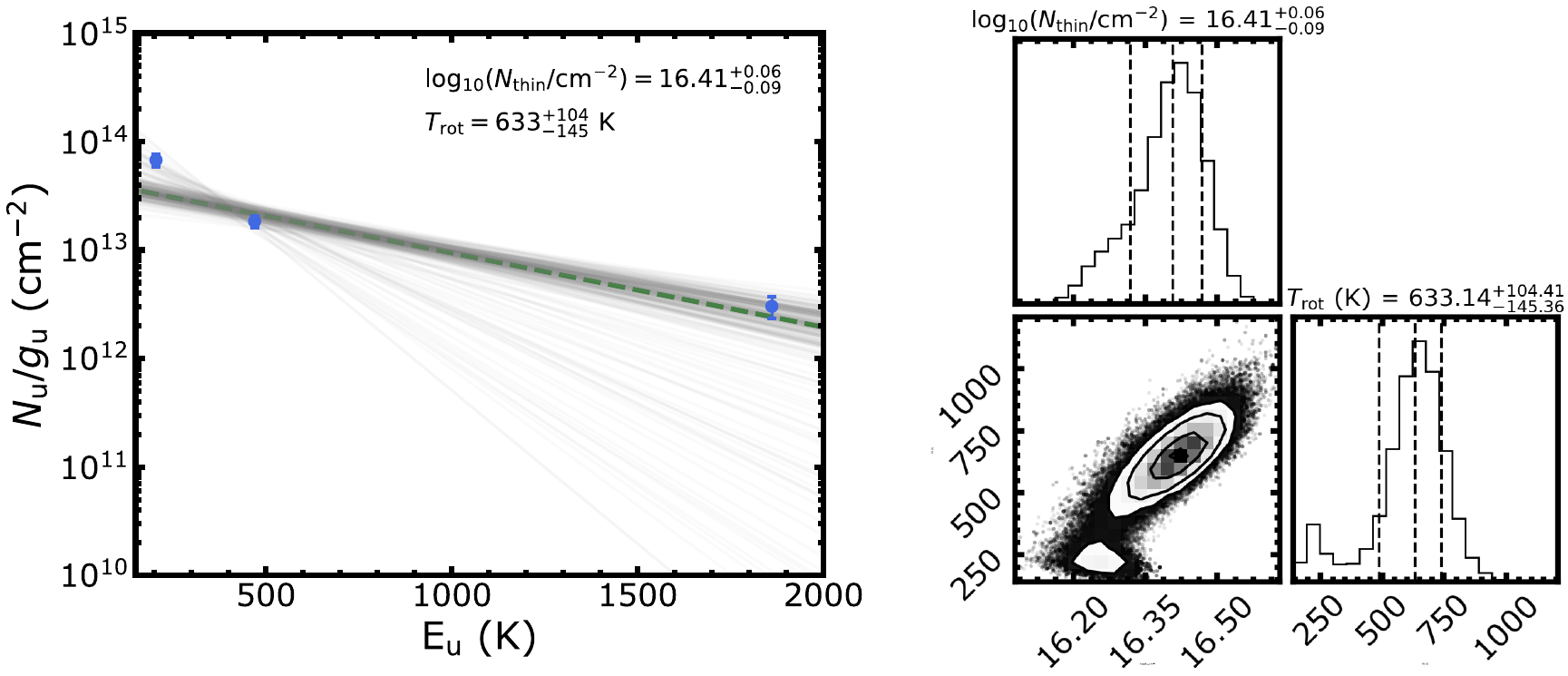}
\end{center}
\caption{Left: rotational diagram of the three water lines, with line fluxes extracted as in the main text. The fit does not lead to a unique solution, indicating that the assumption of uniform temperature is inadequate. Right: posterior distribution of the rotational diagram fit. The dashed lines indicate the 16th, 50th and 84th percentiles of the marginalized posterior distributions.}
\label{fig:T_rot}
\end{figure}

\newpage

\begin{table*}[b]
\centering
\begin{tabular}{lcccccc}
\hline
Transition          & $\nu_0$\,    & $E_{\rm u}$\, & Ch. width & rms$^a$ & Flux & Mask rad$^b$\\
          & (GHz)    &    (K) & (km\,s$^{-1}$) & (mJy\,bm$^{-1}$) & (mJy\,km\,s$^{-1}$) & $('')$\\
\hline
\medskip
p-H$_2$O 3$_{13}$--2$_{20}$  & 183.31009       &  204.7 & 0.8 & 6.90 & $973\pm89$ & 0.70\\
p-H$_2$O 5$_{15}$--4$_{22}$  & 325.15290       &  469.9 & 1.0  & 3.71 & $1332\pm89^c$ & 0.70\\
o-H$_2$O 10$_{29}$--9$_{36}$  & 321.22569       &  1861.3 & 5.0  & 0.96 & $679\pm135$ & 0.28\\
\hline
p-H$_2^{18}$O 5$_{15}$--4$_{22}$  & 322.46517       & 467.9 & 1.0  & 1.75 & $<33.6^d$ & 0.70\\
\hline
\hline
\end{tabular}
\caption{Observed H$_2$O isotopologue transitions. The upper state energies are taken from \cite{JPL}. The notation for the water energy levels in the vibrational ground state is $J_{K_a,K_c}$. Notes: $^a$ Obtained over one single channel. $^b$ Circular radius used to extract the line flux. $^c$~This flux measurement corrects for the absorption identified at $\sim10\,$km\,s$^{-1}$. The flux obtained without accounting for absorption is $929\pm89$\, mJy\,km\,s$^{-1}$. $^d$ $3\sigma$ upper limit.}
\label{tab:water}
\end{table*}

\begin{table*}
\centering
\begin{tabular}{llccccc}
\hline
Program ID          & Lines    & Int. time   & PWV  & Bp/flux cal. & Phase cal. & Max. bl \\
          &     & (min)   & (mm)  &  &  & (m) \\
\hline
\medskip
2017.1.01178.S & p-H$_2$O 3$_{13}$--2$_{20}$  & 46       &   0.2    & J0423-0120 & J0510+1800  & 1398 \\
2017.1.01178.S & p-H$_2$O 5$_{15}$--4$_{22}$  & 31       &   0.4    & J0538-4405 & J0431+1731 & 3637 \\
2017.1.01178.S & o-H$_2$O 10$_{29}$--9$_{36}$  & 33       &   0.5    & J0519-4546  & J0440+1437 & 8547  \\
2022.1.00905.S & p-H$_2$O 5$_{15}$--4$_{22}$  & 100       & 0.3  & J0423-0120    & J0431+1731  & 500 \\
& p-H$_2^{18}$O 5$_{15}$--4$_{22}$  &        &   &     &  &  \\
\hline
\hline
\end{tabular}
\caption{Observations IDs and Execution Blocks properties, including on source integration time, median precipitable water vapour (PWV)  column, bandpass, flux and phase calibrators, and maximum baseline.}
\label{tab:observations}
\end{table*}

\begin{table*}
\centering
\begin{tabular}{lcccccc}
\hline
Transition          &  $g_{\rm u}$   & $A_{\rm ul}$\,  & $Q$\,(100\,K) & $Q$\,(200\,K) & $Q$\,(300\,K) & $Q$\,(400\,K)\\
          &    & (s$^{-1}$)  & (K) &  &  \\
\hline
\medskip
p-H$_2$O 3$_{13}$--2$_{20}$         &  7         & $3.59\times10^{-6}$    & 35.1 &  97.4 & 178.1 & 274.6\\
p-H$_2$O 5$_{15}$--4$_{22}$         &  11        & $1.15\times10^{-5}$    & - & -  & - & - \\
o-H$_2$O 10$_{29}$--9$_{36}$         &  63        & $6.21\times10^{-6}$    & - & -  & - & - \\
\hline
p-H$_2^{18}$O 5$_{15}$--4$_{22}$  &  11        & $1.05\times10^{-5}$    & - & -  & - & - \\
\hline
\hline
\end{tabular}
\caption{Molecular coefficients of the observed H$_2$O isotopologue transitions are from the JPL  database \cite{JPL}, with radiative coefficients from \cite{Barber2006} and the updated partition function $Q$ from the ExoMol database \cite{Polyansky2018}, which well agrees with the one by \cite{Barber2006} in the temperature range explored in this paper. Only some representative values are reported in this table. The degeneracy quantum number of the ortho-state assumes an ortho-to-para ratio of 3. The partition function is computed over all states with the same ortho-to-para ratio.}
\label{tab:water_coeff}
\end{table*}

\clearpage

\section*{Methods}\label{sec2}

\subsubsection*{Observations, data reduction and imaging}
HL Tau was observed in both Band 5 and Band 7 with the ALMA Program 2017.1.01178.S (PI: Humphreys), targeting the two para-water lines at 183.31004\,GHz and 325.15297\,GHz, respectively. HL Tau was also observed in band 7 to target the latter water line with Program 2022.1.00905.S (PI: Facchini), together with the H$_2^{18}$O line at 322.46517\,GHz (see Table~1).  The molecular coefficients for the three transitions are reported in Tables~1 and 3.

Within the 2017.1.01178.S ALMA program, the source was observed in Band 5 on September 21, 2018, for a total integration time of 46\,min, with 43 antennas and baselines ranging between 15\,m and 1.4\,km. The weather conditions during the observation were exceptional, with a median precipitable water vapour (PWV)  column during the observations of $\sim0.19\,$mm. J0423-0120 was used as amplitude and bandpass calibrator, whereas J0510+1800 for phase referencing. The Band 5 spectral setup had 6 spectral windows (spws), five of which targeting different molecular transitions, including the 183\,GHz water line, and a 1.875\,GHz-wide spw for continuum observations at 170.004\,GHz. Within the same program, HL Tau was observed in Band 7 in two spectral setups. The first one on August 12, 2019, with a time on-source of 31\,min, with 48 antennas and baselines ranging between 41\,m and 3.6\,km. With a median PWV column of 0.4\,mm, J0431+1731 was used to cross-calibrate phases, and J0538-4405 for amplitude and spectral response. This first spectral setup consisted of four spws in FDM mode; three of them with a maximum bandwidth of 1.875\,GHz, and one with 1920 244\,kHz channels, targeting the water 325 GHz line. The second spectral setup was used within the same program with an execution block observed on November 24th, 2017. The median PWV was 0.5\,mm. J0519-4546 was used as amplitude and bandpass calibrator, and J0440+1437 as phase calibrator. The observation spent 33\,min on the science target with 49 antennas, with a maximum baseline of 8500\,m. The spectral setup consisted of 4 spws in FDM mode, three of them with maximum bandwidth, and one of them with 1920 244\,kHz channels, targeting the water 321 GHz line.

Finally, new data were taken in October 2022 with a more compact array in Band 7 with Program 2022.1.00905.S, with baselines ranging between 15 and 500\,m, and a total on-source time of 100 min with 41/45 antennas (in two execution blocks). The median PWV column was $\sim0.3$. J0423-0120 was used for flux and bandpass calibration, while J0431+1731 was used for phase referencing. With four spws, one of them was centered on the 325.15\,GHz water line, while another spw targetted the H$_2^{18}$O line at 322.46517\,GHz. 

The Cycle 6 (9) data were calibrated by the ALMA pipeline using \texttt{CASA} v5.4 (v6.4) \cite{casa_software}. For the band 7 data, we combined the data from the two cycles. For both bands, we first self-calibrated each of the three execution blocks in phase, combining all spectral windows after flagging spectral ranges associated to line emission, and combining all scans. Following \cite{Andrews_ea_2018}, we then aligned the data by fitting a Gaussian to the continuum, and shifting the phase-center to the continuum maximum with the {\tt fixvis} and {\tt fixplanet} tasks. We then self-calibrated the two short-baselines (in the case of band 5, only one set of baselines is available) execution blocks in both phase and amplitude, reaching a peak snr of 9950 (a $\sim500\%$ improvement). In the case of band 7, we then combined the long-baseline execution block from the Cycle 6 data, and self-calibrated the data again in both phase and amplitude, reaching a peak snr of 5300 (a $210\%$ improvement). We took particular care with the amplitude calibration, where the gain solution with scan-length intervals greatly improved the data quality. The models for self-calibration were constructed with {\tt CLEAN} with Briggs weighting ({\tt robust}=0.5). The gain solutions were then applied to the full spectral data.

The Band 5 continuum data were imaged with {\tt robust}=0.0. With a synthesized beam of $0.364''\times0.312''$ (PA $-9.5$\,deg), the Band 5 (1.70\,mm) continuum presents an rms of $33\,\mu$Jy, with a peak snr of 2457. The band 7 (0.94\,mm) data were imaged with {\tt robust}=$-1.0$. The data exhibit an rms of $36\,\mu$Jy, with a peak snr of 403, over a synthesized beam of $0.037''\times0.029''$ (PA $1.5$\,deg). A flux density for both images was obtained over an elliptical area with a semimajor axis of $1.3''$, and semiminor axis and position angle (PA) to trace the disk inclination and PA ($46.7$ and $138.0$\, deg, respectively; \cite{HL_Tau_SV_2015}). Without accounting for absolute flux calibration uncertainties, we obtain a flux density of $323.0\pm1.4\,$mJy and $1677.7\pm0.6\,$mJy at 1.70 and 0.94\,mm, respectively, where the uncertainties have been computed as standard deviations of randomly selected masks with the same area of flux density extraction over emission-free regions of the continuum maps.

While the water lines have been imaged with natural weighting (see main text), several additional attempts with a range of uv-tapers were performed to increase the sensitivity to extended emission, but they did not show any feature undetected with natural weighting. Both integrated intensity (moment zero) and intensity weighted velocity (moment one) maps were generated for the water lines. Moment zero maps were computed by integrating channels between $-2$ and $16\,$km\,s$^{-1}$ without any clipping (Figure~1) for the Band 5 line,  between $-6.4$ and $20.6\,$km\,s$^{-1}$ for the 325\,GHz line, and between $-10.4$ and $24.6\,$km\,s$^{-1}$ for the 321\,GHz line. For the two brighter lines, we integrated the moment zero map over a circular area centered over the emission peak and a radius of $0.7''$. We obtain line fluxes of $973\pm89\,$mJy\,km\,s$^{-1}$ and $929\pm89\,$mJy\,km\,s$^{-1}$ (for the 183 and 325\,GHz lines, respectively). Since the 325\,GHz line shows an absorption red-shifted component, we also computed the underlying flux by considering the blue-shifted side only, and multiplying it by a factor of two. The resulting flux is $1332\pm89$\,mJy\,km\,s$^{-1}$. The uncertainties on the line fluxes were computed by bootstrapping over 100 circular apertures in emission free regions of the map, and do not account for absolute flux calibration uncertainties. The same operation was applied to the 321\,GHz line, using a smaller $0.28''$ radius extraction area. The resulting flux is $679\pm135\,$mJy\,km\,s$^{-1}$. The flux is much lower than the tentative detection by \cite{Kristensen_ea_2016} with the Submillimeter Array (SMA), where however the line is shifted by 30\,km\,s$^{-1}$ from the systemic velocity, indicating that the proposed emission may be originating from a large scale flow which we filter out in our high resolution data. For the 183 and 325\,GHz lines, intensity weighted velocity maps were generated using the \texttt{bettermoments} package \cite{bettermoments} after applying $4$ and $3$\,$\sigma$ clipping to individual channels, respectively. 

The non detection of the H$_2^{18}$O 322\,GHz line is shown in Extended Data Figure~5.

\subsubsection*{Fitting of the 183\,GHz line spectrum}

Given the high snr of the 183\,GHz line, we fitted its spectrum by analytically computing model spectra of a geometrically thin Keplerian disk, similarly to \cite{Zagaria2023}. To do so, we assumed that the peak brightness temperature of the line decreases with radius as a power-law:

\begin{equation}
    T(R) = T_0 \left(\frac{R}{10\,{\rm au}} \right)^{-q}.
\end{equation}
For a given radius R, and azimuth $\phi$, we assumed that the line intensity follows a Gaussian distribution velocity:
\begin{equation}
I(R,\phi,v)=\frac{B_{\nu}(T)}{d^2} {\rm exp}\left(-\frac{\mu m_{\rm H}(v-v_{\rm K,proj})^2}{2k_{\rm B}T}\right),
\end{equation}
where $B_{\nu}(T)$ is the Planck function at temperature T, $k_{\rm B}$ is the Boltzmann constant, $d$ is the distance of HL Tau (140\,pc), $\mu m_{\rm H}$ is the mass of the water molecule, and $v_{\rm K,proj}$ can be written as follows:

\begin{equation}
    v_{\rm K,proj} = \left( \frac{GM_\odot}{R} \right)^{1/2}\sin{i}\cos{\phi},
\end{equation}
where $i$ is the source inclination ($46.7\deg$). We considered an optically thick limit when assuming a thermal broadening with a kinetic temperature equalling the brightness temperature. The flux density of the line can then be computed at every velocity $v$ by integrating across the whole disk:
\begin{equation}
F(v) = \cos{i}\int_0^{2\pi}\int_{R_{\rm in}}^{R_{\rm out}}I(R,\phi,v)RdRd\phi,
\end{equation}
where $\cos{i}$ accounts for the geometrical projection on the sky. We fixed $R_{\rm in}$ to $0.1\,$ au (but the model is not sensitive to this value for reasonably small radii), and sampled the disk with 150 points in radius, and 550 in azimuth. We then convolved the models with a Gaussian kernel with the same channel width as in the data, and sampled them at the same velocities. We kept three free parameters in the fitting procedure: $T_0$, $q$ and $R_{\rm out}$. We fitted the spectrum shown in Figure~2 with the {\tt emcee} package, using flat priors on the three free parameters: [10,1500]\,K, [0,3], [2,200]\,au, respectively. We used 30 walkers, 1000 steps of burn-in, and 1000 additional steps to sample the posterior distribution. Extended Data Figure~6 shows the best fit model, and 100 random draws extracted from the posterior distribution. While the fit does not manage to constrain the outer radius of the emission, we obtain $T_0=287_{-154}^{+180}$, and $q=0.92^{+0.30}_{-0.47}$. The fit highlights a negative brightness temperature gradient in the radial profile, as seen in the reconstructed integrated intensity profile (see Figure~3), and as hinted by the rotational diagram shown in Figure~4.

\subsubsection*{Parametric fit of the integrated intensity profile in the visibility plane}

In order to extract the radial extent of the 183\,GHz line, we performed a parametric fit of its integrated intensity radial profile in the visibility plane, by exploiting the {\tt galario}  package. After averaging the continuum-subtracted visibilities in the same spectral range used to compute the moment zero map, we fitted the visibility data by Fourier transforming a projected integrated intensity profile in the same $uv$-points sampled during the observations. We modelled the radial profile with a simple Gaussian prescription:
\begin{equation}
J(R) = J_0\,{\rm exp}\left(-\frac{R^2}{2\sigma_{\rm G}^2}\right),
\end{equation}
where we considered four free parameters: $J_0$, $\sigma_{\rm G}$, and the disk center ($\Delta$RA and $\Delta$Dec). We fixed the inclination and position angle to the ones obtained from high resolution continuum imaging \cite{HL_Tau_SV_2015}. The fit was performed with the {\tt emcee} package, where the $J_0$ parameter was sampled in log-space. We used the following flat priors on the parameters: $\log_{10}(J_0/{\rm steradian})\in[1,20]$, $\sigma_{\rm G}\in[0,1.5]''$, $\Delta{\rm RA}\in[-0.4,0.4]''$, $\Delta{\rm Dec}\in[-0.4,0.4]''$. The posterior distribution was sampled with 50 walkers and 1000 steps, after 1000 steps of burn-in. The MCMC exploration of the posterior space well converges, as shown in Extended Data Figure~7.

\subsubsection*{Rotational diagram and optical depth constraints}

To compute the rotational diagram of the water molecule, we use the same approach as in e.g., \cite{Loomis18,Facchini21}. In the optically thin assumption, we can compute the column density $N_{\rm thin}$ and the rotational temperature $T_{\rm rot}$ by measuring the integrated flux $S_{\!\nu} \Delta v$:

\begin{equation}
\label{eq:column}
    N_{\rm thin} = \frac{4\pi}{A_{\rm ul} h c} \frac{S_{\!\nu} \Delta v}{\Omega} \frac{Q(T_{\rm rot})}{g_{\rm u}} \, \exp\left(\frac{E_{\rm u}}{T_{\rm rot}}\right),
\end{equation}
where $\Omega$ is the solid angle used for flux extraction (see previous section), $A_{\rm ul}$ is the Einstein coefficient of the considered transition, $h$ and $c$ are the Planck constant and the speed of light in vacuum, $Q$ is the partition function, $E_{\rm u}$ is the upper state energy (in K) and 
$g_{\rm u}$ the upper state degeneracy. Using the relation between $N_{\rm u,thin}$ and $N_{\rm thin}$, the same equation can be written in logarithmic form \cite{Goldsmith99}:
\begin{equation}
\ln {\frac{N_{\rm u,thin}}{g_{\rm u}}} = \ln{N_{\rm thin}} - \ln{Q(T_{\rm rot})} - E_{\rm u}/T_{\rm rot}.
\end{equation}
We performed a linear regression using the {\tt emcee} sampler \cite{Foreman-Mackey+13} in the $[\ln {(N_{\rm u,thin}/g_{\rm u})},\, E_{\rm u}]$ space to extract $N_{\rm thin}$ and $T_{\rm rot}$.  We used the molecular coefficients reported in Table~3. In the fitter, the partition function was determined with a cubic spline interpolation across the rotational temperatures listed in \cite{Polyansky2018}. The same approach was used for the rotation diagram analysis with six to eight transitions of o- and p-water in evolved stars \cite{Baudry2023}. In the Monte Carlo Markov Chain (MCMC) sampling, we used 128 walkers, and 2000 steps (after 1000 steps of burn-in). While Figure~4 shows the result and the marginalized posterior distributions of the fit of all three lines, Extended Data Figure~8 portrays the individual fits on the two colder and warmer lines, respectively.

To compute the upper limit on the optical depth of the 325\,GHz water line, we exploited the non-detection of the H$_2^{18}$O line, which has almost identical molecular coefficients. Assuming that the column density of the main isopotologue line ($\tau_{\rm{H}_{2}\rm{O}}$) is equal to $530\times\tau_{\rm{H}_{2}^{18}\rm{O}}$ \cite{McKeegan11}, we can write:

\begin{equation}
\frac{(S_{\!\nu} \Delta v)_{\rm{H}_{2}\rm{O}}}{(S_{\!\nu} \Delta v)_{\rm{H}_{2}^{18}\rm{O}}} \approx \frac{1-e^{-\tau_{\rm{H}_{2}\rm{O}}}}{1-e^{-\tau_{\rm{H}_{2}\rm{O}}/530}}.
\end{equation}
Using the $3\sigma$ upper limit on the H$_2^{18}$O transition, and the measured flux of the H$_2$O line (after correcting for absorption, since the absorbing column of H$_2^{18}$O will have the same scaling factor of 530), a numerical solution of the equations leads to $\tau_{\rm{H}_{2}\rm{O}}<14$, as reported in the main text.

\newpage

\section*{Data availability} All the ALMA data are publicly available on the ALMA archive (https://almascience.nrao.edu/aq/), with program IDs \#2017.1.01178.S and \#2022.1.00905.S.

\section*{Code availability} The python packages used in the data analysis are all publicly available. The calibration and fitting scripts can be obtained by SF upon reasonable requests.

\section*{Correspondence}
Correspondence and requests for materials should be addressed to Stefano Facchini, stefano.facchini@unimi.it.

\section*{Acknowledgments}

We are thankful to the three reviewers for their thorough and insightful reports which helped improve the clarity of the paper. This paper makes use of the following ALMA data: ADS/JAO.ALMA\#2017.1.01178.S, ADS/JAO.ALMA\#2022.1.00905.S. ALMA is a partnership of ESO (representing its member states), NSF (USA) and NINS (Japan), together with NRC (Canada), MOST and ASIAA (Taiwan), and KASI (Republic of Korea), in cooperation with the Republic of Chile. The Joint ALMA Observatory is operated by ESO, AUI/NRAO and NAOJ. S.F. is funded by the European Union (ERC, UNVEIL, 101076613). Views and opinions expressed are however those of the author(s) only and do not necessarily reflect those of the European Union or the European Research Council. Neither the European Union nor the granting authority can be held responsible for them. S.F. acknowledges financial contribution from PRIN-MUR POPS 2022YP5ACE. L.T. acknowledges funding from Progetti Premiali 2012 – iALMA (CUP C52I13000140001), Deutsche Forschungs-gemeinschaft (DFG, German Research Foundation) - Ref no. 325594231 FOR 2634/1 TE 1024/1-1, European Union’s Horizon 2020 research and innovation programme under the Marie Sklodowska- Curie grant agreement No 823823 (DUSTBUSTERS) and the European Research Council (ERC) via the ERC Synergy Grant ECOGAL (grant 855130). M. V. D. acknowledges support from Wallonie-Bruxelles International (Belgium) through its grant ``Stage en Organisation International'' and the French-speaking Community of Belgium through its FRIA grant.

\section*{Author Contributions Statement} S.F. and E.H. led the ALMA observing proposals. S.F. calibrated the data and performed the data analysis. S.F. and L.T. wrote the manuscript. All co-authors provided input on the manuscript. 

\section*{Competing Interests Statement}
The authors declare no competing interests.

\newpage

\section*{Extended data}

\begin{figure*}[b]
\begin{center}
\includegraphics[width=0.48\columnwidth]{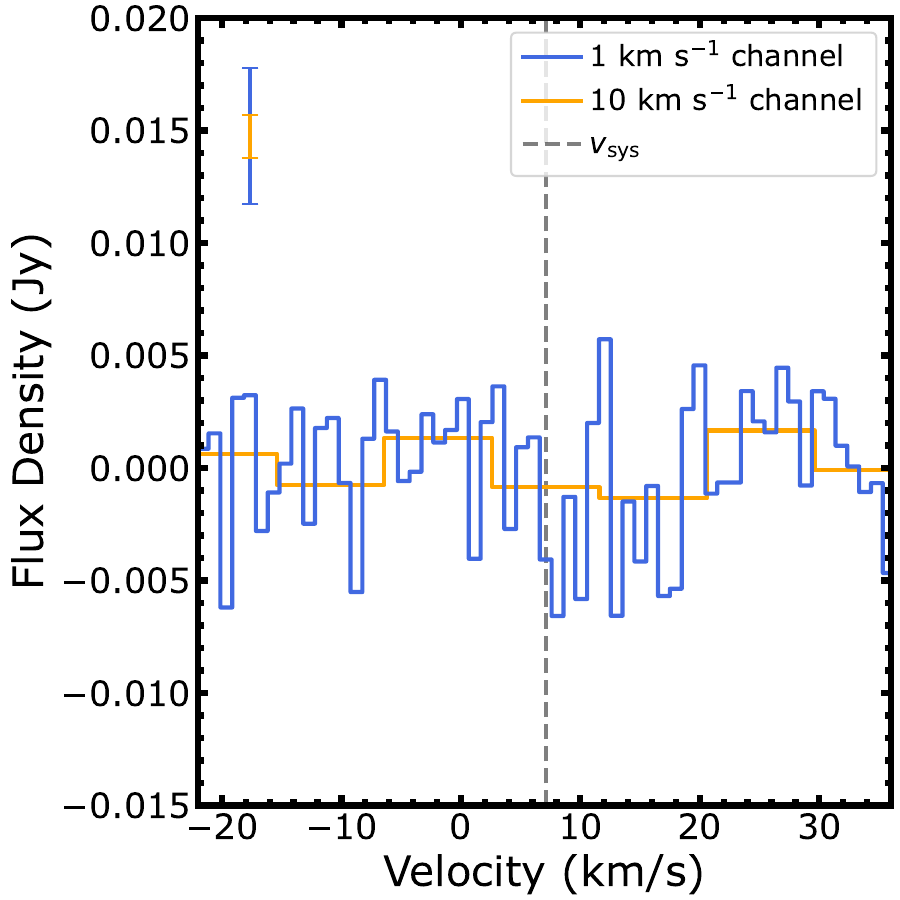}
\end{center}
\caption{Spectrum of the 322\,GHz H$_2^{18}$O line, extracted over a circle with radius of $0.7''$ centered on the continuum peak, as for Fig.~2. Two spectra are shown, from channel maps with two different channel widths. The grey dashed line indicates the systemic velocity of $7.1\,$km\,s$^{-1}$ \cite{Garufi_ea_2021,Garufi_ea_2022}. In the top left 2$\sigma$ scale bars are reported for reference.}
\label{fig:spectrum_h218o}
\end{figure*}

\begin{figure*}
\begin{center}
\includegraphics[width=\columnwidth]{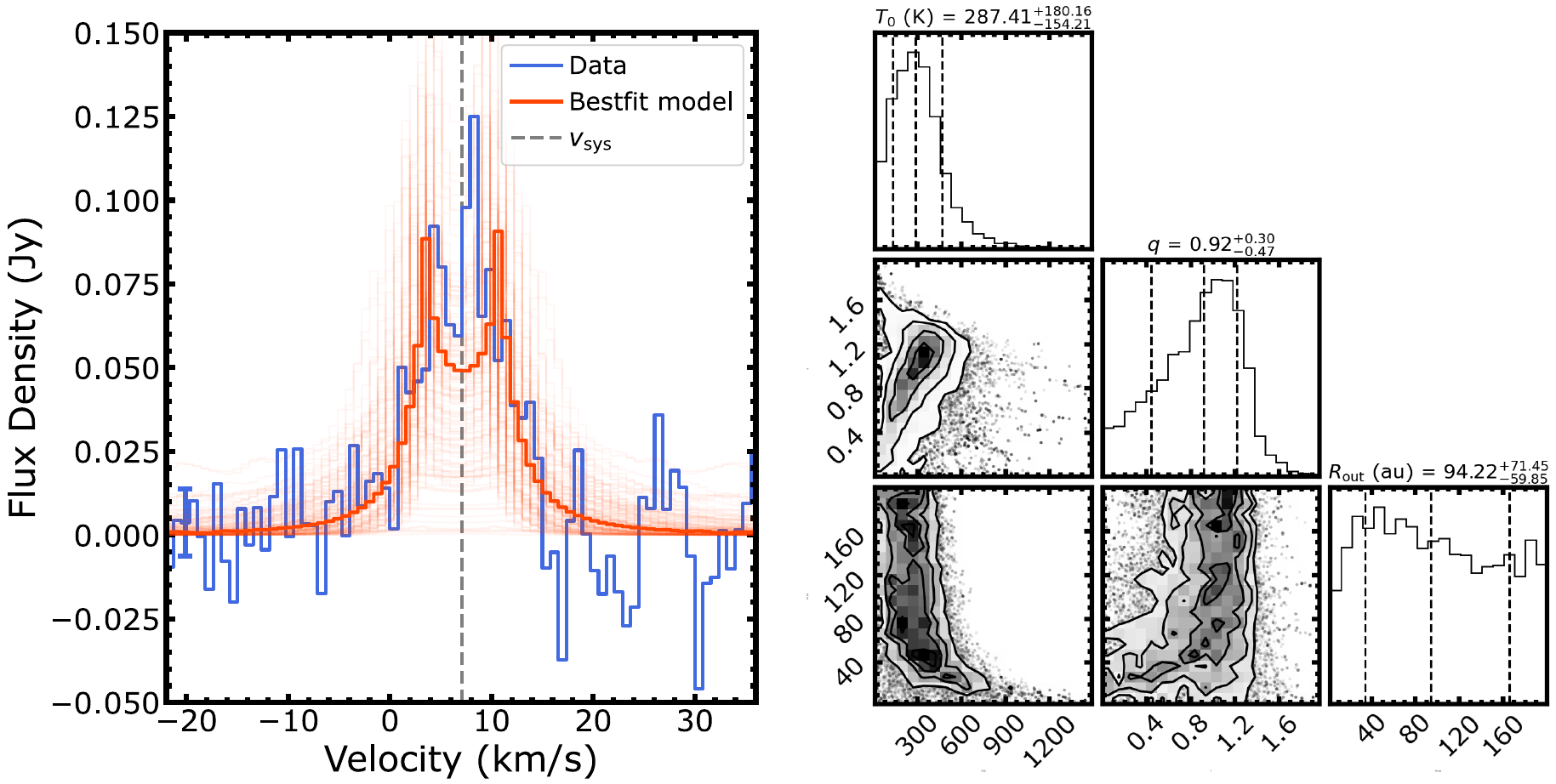}
\end{center}
\caption{Left: spectrum of the 183\,GHz H$_2$O line, as in Fig.~2. The uncertainty associated to each data point is shown in the left part of the spectrum as an errorbar. A random sampling of 100 profiles from the posterior distribution of the MCMC fit is shown, with the dark red line indicating the bestfit model. Right: marginalized posterior distribution of the fitted parameters, where no constraint can be retrieved on $R_{\rm out}$ from this analysis.}
\label{fig:spectrum_b5_fit}
\end{figure*}

\begin{figure}
\begin{center}
\includegraphics[width=0.7\columnwidth]{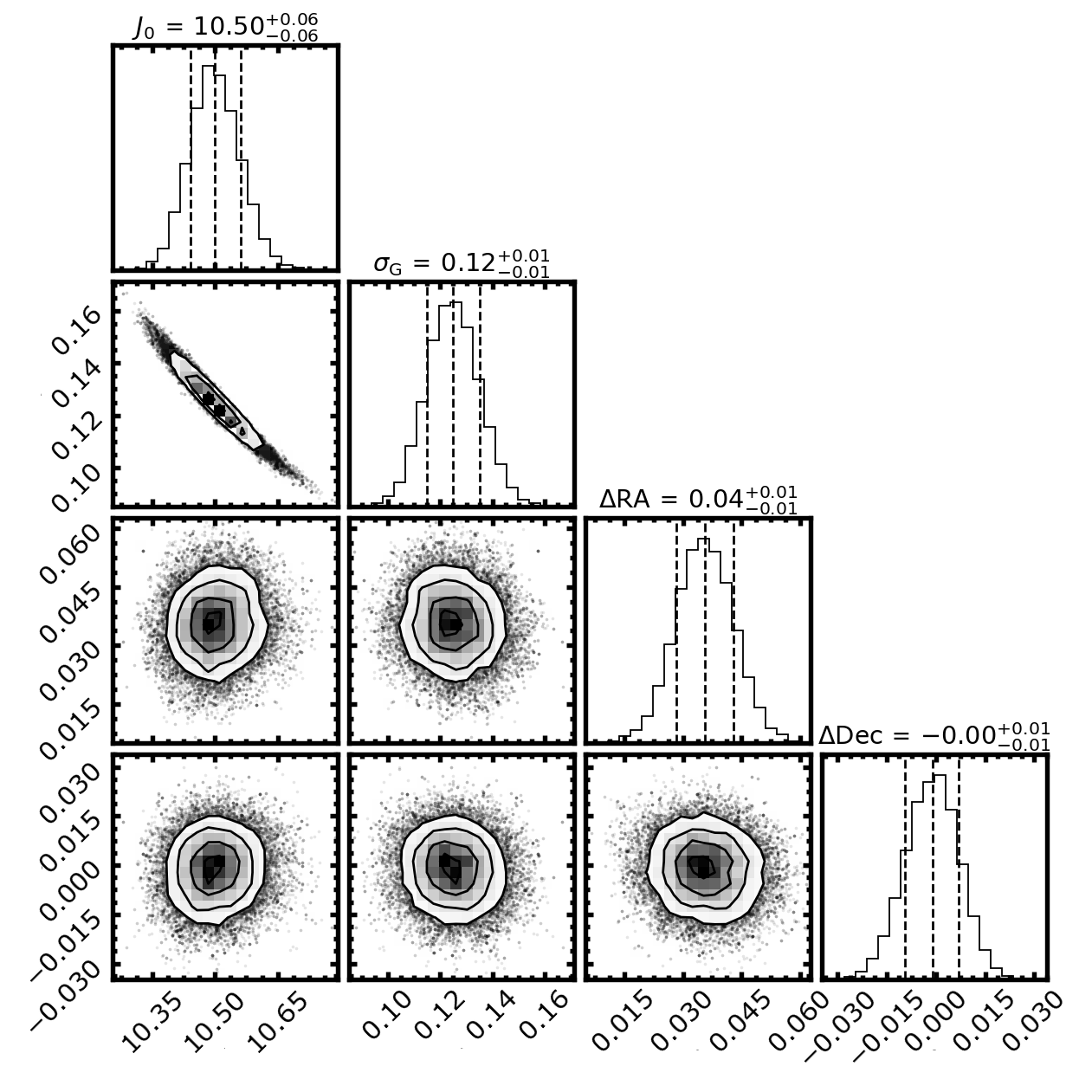}
\end{center}
\caption{Marginalized posterior distribution of the {\tt galario} fit of the visibility points of the integrated intensity of the 183\,GHz line (see Fig.~3).}
\label{fig:triangle_b5}
\end{figure}

\begin{figure*}
\begin{center}
\includegraphics[width=\columnwidth]{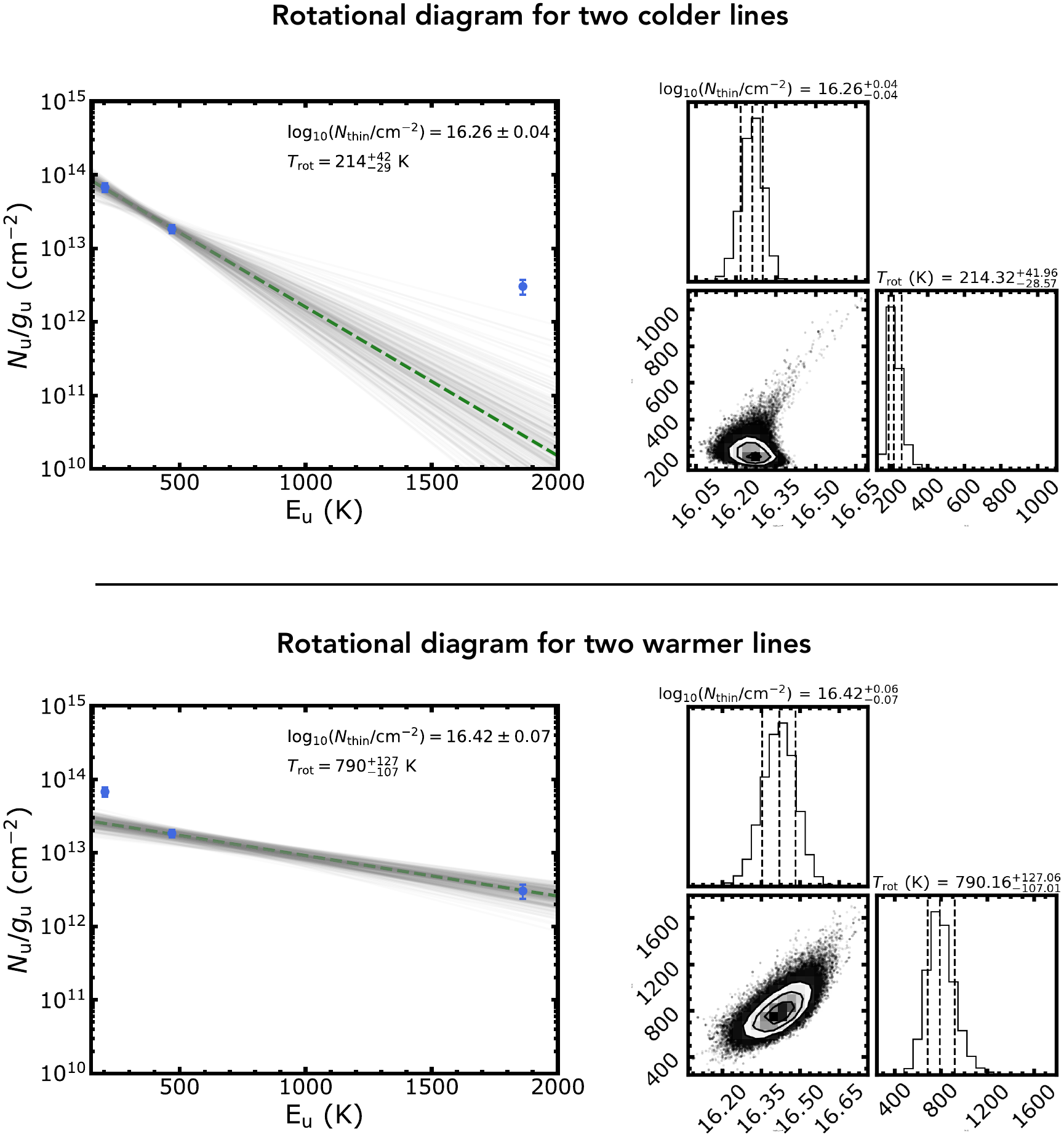}
\end{center}
\caption{Rotational diagrams and marginalized posterior distribution of the MCMC exploration for individual fits on the two lowest and two highest energy lines, respectively.}
\label{fig:triangle_trot}
\end{figure*}

\clearpage

\bibliography{scibib}

\bibliographystyle{sn-nature.bst}

\end{document}